\documentclass[prl,showpacs,showkeys, twocolumn]{revtex4}

\usepackage{amsmath}
\usepackage{amsfonts}            
\usepackage{amssymb}

\usepackage{mathrsfs}

\usepackage{eufrak}

\usepackage{graphicx}
\usepackage{epstopdf}

\usepackage{dcolumn}
\usepackage{bm}

\makeatletter
\def\@dotsep{4.5}
\makeatother

\begin{document}


\title{Supersymmetric factorization yields exact solutions \\ to the molecular Stark effect problem for ``stretched'' states}

\author{Mikhail Lemeshko}

\email{mikhail.lemeshko@gmail.com}

\affiliation{%
Fritz-Haber-Institut der Max-Planck-Gesellschaft, Faradayweg 4-6, D-14195 Berlin, Germany
}%

\author{Mustafa Mustafa}

\email{mustafa@purdue.edu}


\author{Sabre Kais}

\email{kais@purdue.edu}

\affiliation{%
Departments of Chemistry and Physics,  Purdue University, West Lafayette, Indiana 47907, USA
}%

\author{Bretislav Friedrich}

\email{brich@fhi-berlin.mpg.de}

\affiliation{%
Fritz-Haber-Institut der Max-Planck-Gesellschaft, Faradayweg 4-6, D-14195 Berlin, Germany~}%

\date{\today}

\begin{abstract}

By invoking supersymmetry, we found a condition under which the Stark effect problem for a polar and polarizable molecule subject to nonresonant electric fields becomes exactly solvable for the $|\tilde{J}=m,m\rangle$ family of ``stretched'' states. The analytic expressions for the wavefunction and eigenenergy and other expectation values allow to readily reverse-engineer the problem of finding the values of the interaction parameters required for creating quantum states with preordained characteristics. The method also allows to construct families of isospectral potentials, realizable with combined fields.

\end{abstract}

\pacs{32.60.+i, 33.90.+h, 33.15.Kr,11.30.Pb, 37.10.Vz, 03.65.-w, 03.65.Ge}
\keywords{orientation, alignment, Stark effect, combined fields, induced-dipole interaction, supersymmetry in quantum mechanics, shape-invariance, Schr\"odinger equation, exact solvability, analytic wavefunction} 

\maketitle

In 1983, Gendenshtein demonstrated that Schr\"odinger's equation is exactly solvable
if the potential and its superpartner exhibit shape-invariance~\cite{Gendenshtein83}. Whereas a supersymmetric Hamiltonian can be constructed for any potential whose ground-state wavefunction is analytic, shape invariance only exists for supersymmetric potentials that are interconvertible by a change of a parameter other than the integration variable itself~\cite{BougieMallowPRL10, KaisJCP89}. Herein, we make use of the methods of supersymmetric quantum mechanics to arrive at exact wavefunctions and other eigenproperties of molecules subject to nonresonant electric fields  in closed form.

In our previous work on the molecular Stark effect, we showed that for polar molecules, combined collinear electric and nonresonant radiative fields can synergetically produce spatially oriented pendular states, in which the molecular axis librates over a limited angular range about the common field direction~\cite{FriHerJCP99,HaerteltFriedrichJCP08}. These directional states comprise hybrids of the field-free rotational states $|J,m\rangle$, with a range of $J$ values but a fixed value of $m$, which remains a good quantum number by virtue of the azimuthal symmetry about the fields. This has proved an effective and versatile means to produce oriented molecules for applications ranging from molecule optics and spectroscopy to chemistry and surface science~\cite{StapelfeldtSeidemanRMP03, KremsPCCP08, VattuoneProgSurfSci10}. However, the eigenproperties of the Stark states in question could only be found numerically, typically by diagonalizing a truncated Hamiltonian matrix. This renders all the treatments of molecular processes in fields, such as cold collisions or collective behavior of ultracold polar gases, analytically unsolvable~\cite{KreStwFrieColdMolecules}. Here we show that supersymmetric factorization of the Hamiltonian yields exact wavefunctions $|\tilde{J}=m,m;\omega,\Delta \omega\rangle$ in closed form for a particular ratio of the parameters $\omega$ and $\Delta \omega$ that determine the interaction strengths of the molecules with the static and radiative fields, respectively. We found that, in semiclassical terms, this ratio originates in the integrability of the differential equation for the system's action. We also found that in the exactly solvable field-free and strong-field limits the supersymmetric problem indeed exhibits shape-invariance. 

We consider a $^1\Sigma$ molecule with a rotational constant $B$, a permanent dipole moment $\mu$ along the internuclear axis, and polarizability components $\alpha_\parallel$ and $\alpha_\perp$ parallel and perpendicular to the internuclear axis. The molecule is subjected to an electrostatic field $\boldsymbol{\varepsilon}$ combined with a nonresonant laser field of intensity $I$, whose linear polarization is collinear with $\boldsymbol{\varepsilon}$. With energy expressed in terms of $B$, the Hamiltonian takes the dimensionless form~\cite{FriHerJCP99},
\begin{equation}
	\label{HcombinedFields}
	H =  \mathbf{J}^2  - \omega \cos \theta - \left( \Delta\omega \cos^2 \theta + \omega_\perp   \right),
\end{equation}
with the dimensionless interaction parameters $\omega \equiv \mu \varepsilon/B$,  $\Delta \omega \equiv \omega _{||}-\omega _{\bot }$, and $\omega _{||,\bot } \equiv 2\pi \alpha_{||,\bot }I/(Bc)$. 

The common direction of the collinear electrostatic and linearly polarized radiative fields defines an axis of cylindrical symmetry, chosen to be the space-fixed axis $Z$. The projection, $m$, of the angular momentum $\mathbf{J}$ on $Z$ is then a good quantum number while $J$ is not. However, one can use the value of $J$ of the field-free rotational state, $Y_{J, m} (\theta, \phi)$, that adiabatically correlates with the hybrid state as a label, designated by $\tilde{J}$, so that $|\tilde{J}, m;\omega, \Delta \omega\rangle \to Y_{J, m}$ for $\omega, \Delta \omega \to 0$. For arbitrary values of the interaction parameters $\omega$ and $\Delta \omega$, the solution to the Schr\"odinger equation with Hamiltonian~(\ref{HcombinedFields}) is an infinite coherent superposition of the field-free rotor wavefunctions, whose expansion coefficients can be obtained by truncating the series and diagonalizing the Hamiltonian in the resulting finite basis set. 

The axial symmetry of the problem allows to separate angular variables and express the dependence on the azimuthal angle $\phi$ via the good quantum number $m$. The Schr\"odinger equation for Hamiltonian~(\ref{HcombinedFields}) then becomes

\begin{multline}
	\label{SEcombinedfields}
	\Biggl[ - \frac{1}{\sin\theta} \frac{d}{d \theta} \left( \sin\theta  \frac{d}{d \theta}  \right)  + \frac{m^2}{\sin^2 \theta} \\ 
	- \omega \cos \theta - \Delta \omega \cos^2 \theta  \Biggr]   \psi (\theta) = E \psi (\theta).
\end{multline}

We note that all rotational levels are uniformly shifted by $\omega_\perp$. In what follows we use  $E = E_{\mu,\alpha} + \omega_\perp$ instead of the `true' molecular energy, $E_{\mu,\alpha}$. Moreover, since the Stark effect does not depend on the sign of $m$, we  define the projection of the angular momentum on $Z$ as a positive quantity, $m \equiv |m|$. 

By means of the substitution, $\psi (\theta) = f (\theta) (\sin  \theta)^{-\frac{1}{2}}$, eq.~(\ref{SEcombinedfields}) can be transformed to a one-dimensional form~\cite{InfeldHullRMP51},

\begin{equation}
	\label{SEtransformed}
	\Biggl [ -\frac{d^2}{d \theta^2} + \frac{m^2 - \frac{1}{4}}{\sin^2\theta} - \omega \cos\theta - \Delta \omega \cos^2\theta  - \frac{1}{4}   \Biggr ] f (\theta) = E f (\theta),
\end{equation}
which will be shown to play the role of one of the requisite superpartner equations leading to ground state energy $E=E_0$.

In what follows, we invoke supersymmetry to find analytic solutions to eq. (\ref{SEtransformed}) and subsequently to eq. (\ref{SEcombinedfields}). Supersymmetry makes use of the first-order differential operators, $A^\pm \equiv \mp \frac{d}{d\theta} + W(\theta)$, with $W(\theta)$ being the superpotential. The superpartner Hamiltonians are defined by
\begin{equation}
	\label{HminusFactor}
	H_\mp = A^\pm A^\mp =  -\frac{d^2}{d\theta^2} +  V_\mp^\text{(1D)}(\theta),
\end{equation}
with the one-dimensional partner potentials $V_\pm^\text{(1D)} (\theta) \equiv W^2(\theta) \pm W'(\theta)$. The superpartner Hamiltonians have the same energy spectra except for the ground state, i.e., $E_{n+1}^- = E_n^+$ and $E_0^- = 0$. If the eigenfunctions of one of the partner Hamiltonians $H_\mp$ are known, the eigenfunctions of the other can be obtained analytically via the intertwining relations, $\psi_{n-1}^+ \sim A^- \psi_n^-$ and $\psi_{n}^- \sim A^+ \psi_{n-1}^+$~\cite{CooperPhysRep95, SukumarJPA85}.

For a molecule in the combined fields, we assume the superpotential to have the form $W(\theta) = a \cot \theta + q(\theta)$, where the first term corresponds to the field-free rotor~\cite{InfeldHullRMP51, DuttAJP97}. By identifying the effective potential in eq.~(\ref{SEtransformed}) with $V_-^\text{(1D)} (\theta)$, the constant $a$ and the function $q(\theta)$ can be determined, leading to the following expression for $W(\theta)$,
\begin{equation}
	\label{W}
	W(\theta) =  - \left( m + \frac{1}{2} \right) \cot \theta + \beta \sin \theta,
\end{equation}
and the corresponding SUSY partner potentials,
\begin{equation}
	\label{Vminus1D}
	V_-^\text{(1D)}(\theta) =   \frac{m^2 - \frac{1}{4}}{\sin^2 \theta} - 2\beta(m+1) \cos\theta - \beta^2\cos^2\theta - E_0 -\frac{1}{4} 
\end{equation}
\begin{equation}
	\label{Vplus1D}
	V_+^\text{(1D)}(\theta) =    \frac{(m+1)^2 - \frac{1}{4}}{\sin^2 \theta} - 2\beta m \cos\theta - \beta^2\cos^2\theta - E_0 -\frac{1}{4}
\end{equation}
with $E_0  = m(m+1) - \beta^2$. The strengths of the combined fields are connected with $\beta$ via the following expression,
\begin{equation}
	\label{OmegaDeltaOmega}
	\Delta \omega = \frac{\omega^2}{4 (m+1)^2 } = \beta^2.
\end{equation}

The ground state wavefunction $f_0^-(\theta)$ can be obtained from superpotential (\ref{W}) in closed form~\cite{CooperPhysRep95},
\begin{equation}
	\label{fviaW}
	f_0^-(\theta)  = N (-1)^m (\sin\theta)^{(m+1/2)}~e^{\beta \cos \theta};
\end{equation}
the normalization constant $N$ can be expressed analytically via the hypergeometric functions~\cite{LemMusKaisFriLong, AbramowitzStegun}. The phase factor $(-1)^m$ leads to the correct asymptotic behavior of the wavefunction $f_0^-(\theta) (\sin  \theta)^{-\frac{1}{2}}$ which, for $\beta = 0$, reduces to the ground state wavefunction of a rigid rotor with $J=m$, $Y_{m,m} (\theta, 0)$.
Since the ground-state wavefunction (\ref{fviaW}) is normalizable and obeys the annihilation condition $A^- f_0^- = 0$, the supersymmetry obtained is unbroken~\cite{CooperPhysRep95, WittenNuclPhysB82, CooperFreedmanAnnPhys83}.


\begin{figure}[htbp]
\includegraphics[width=8.5cm]{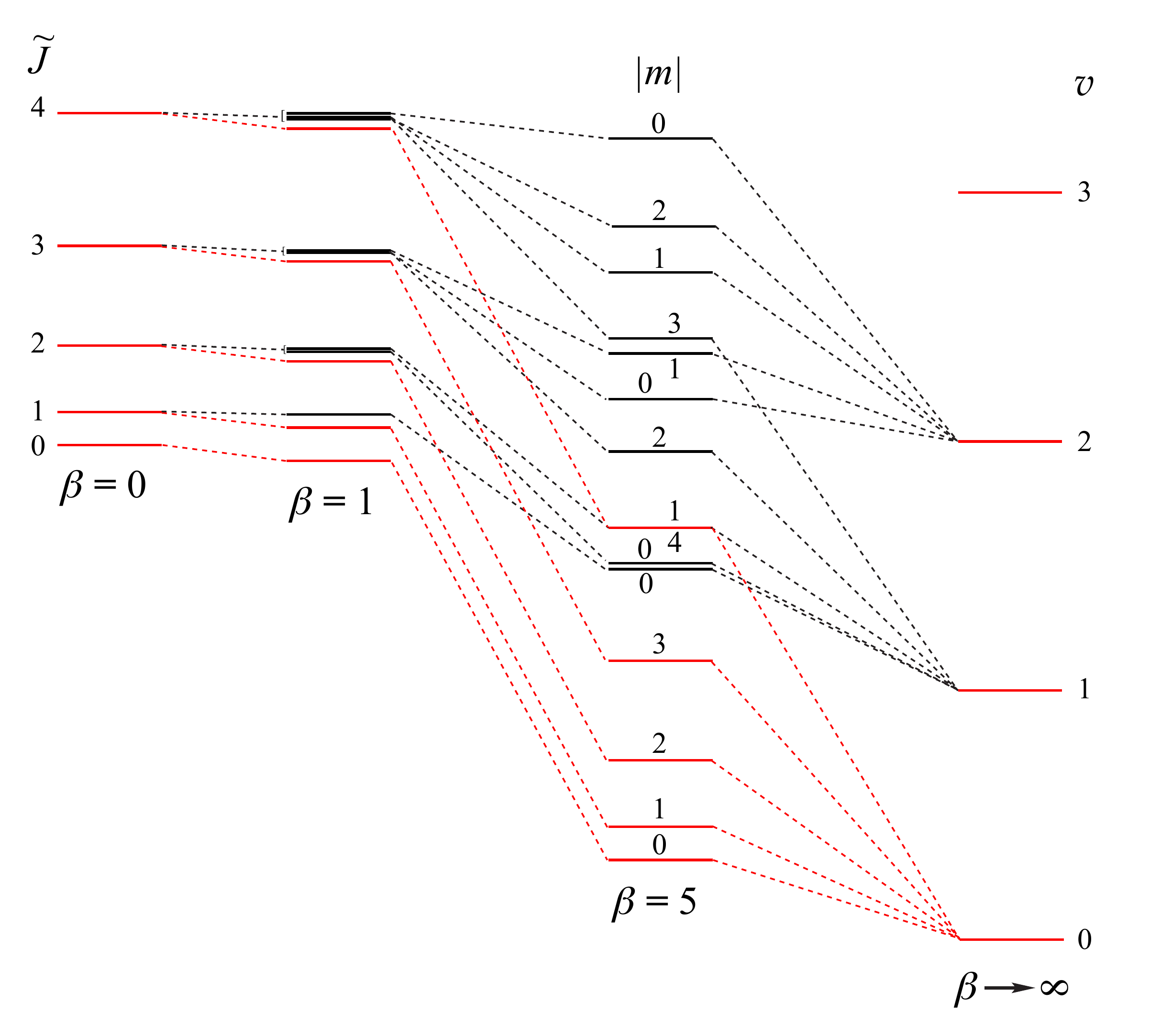}
\caption{\label{fig:corr} Correlation diagram for a molecule in combined fields
as a function of $\beta$. Eigenstates that can be obtained in closed form are shown in
red. The equidistant energy levels in the strong-field limit, $\beta
\to \infty$, are labeled by the 3D librator quantum number $v$. See text.}
\end{figure}

The 3D potentials corresponding to a molecule interacting with the combined fields, can be obtained by comparing the $V_+^\text{(1D)}$ and $V_-^\text{(1D)}$, eqs.~(\ref{Vminus1D}),~(\ref{Vplus1D}), with eqs.~(\ref{SEcombinedfields}) and (\ref{SEtransformed}). The detailed analysis of the 1D and 3D supersymmetric partner potentials is presented elsewhere~\cite{LemMusKaisFriLong}.


Eq.~(\ref{Vplus1D}) coincides with the potential for a rigid rotor in the combined fields whose projection quantum number is $m+1$ and whose interaction strengths are related by
\begin{equation}
	\label{OmegaDelatOmega2}
	\Delta \omega = \frac{\omega^2}{4 m^2 } = \beta^2.
\end{equation}
Hence, given a value of $\beta$, the Hamiltonian of a molecule with a projection  $m$ of the angular momentum on the combined fields whose interaction parameters are related via eq. (\ref{OmegaDeltaOmega}) has the same set of eigenvalues as a molecule with a projection $m+1$ on the combined fields whose interaction parameters are related by eq.~(\ref{OmegaDelatOmega2}).


Unfortunately, the partner potentials (\ref{Vminus1D}) and (\ref{Vplus1D}) are not  shape invariant and so eq.~(\ref{SEtransformed}) for a molecule in the combined fields is not, in general, exactly solvable~\cite{FriHerJCP99, AbramowitzStegun}. We also note that none of the known shape-invariant superpotentials listed, e.g., in refs.~\cite{DuttKhareSukhatmeAJP88, BougieMallowPRL10} leads to exactly solvable partner Hamiltonians that can be experimentally implemented for molecules in nonresonant fields. 

Figure~\ref{fig:corr} shows the energy levels of a molecule in combined fields for different values of the field-strength parameter $\beta$. In the weak-field limit, $\beta \to 0$, the energy levels approach those of a free-rotor, which is solvable exactly for all the eigenstates. For nonzero but weak fields, $\beta=1$, the levels split into $\tilde{J}+1$ components due to the Stark effect. In this case, the SUSY partner Hamiltonians are not shape-invariant, and the problem is analytically solvable only for the ``stretched states,'' with $\tilde{J} = m$. With increasing interaction strength, the energies of the stretched states come closer to one another and, in the strong-field limit, $\beta \to \infty$, coalesce into the ground state level of the 3D harmonic librator. In the strong-field limit, the supersymmetric problem becomes shape-invariant again, and is exactly solvable for all eigenstates in closed form; the equidistant levels (labeled by the quantum number $v$) are infinitely degenerate and separated by an energy difference of $(2\omega + 4 \Delta \omega)^{1/2}$. The weak-field and strong-field limits are described in detail in a forthcoming paper~\cite{LemMusKaisFriLong}.

\begin{figure}
\includegraphics[width=7.2cm]{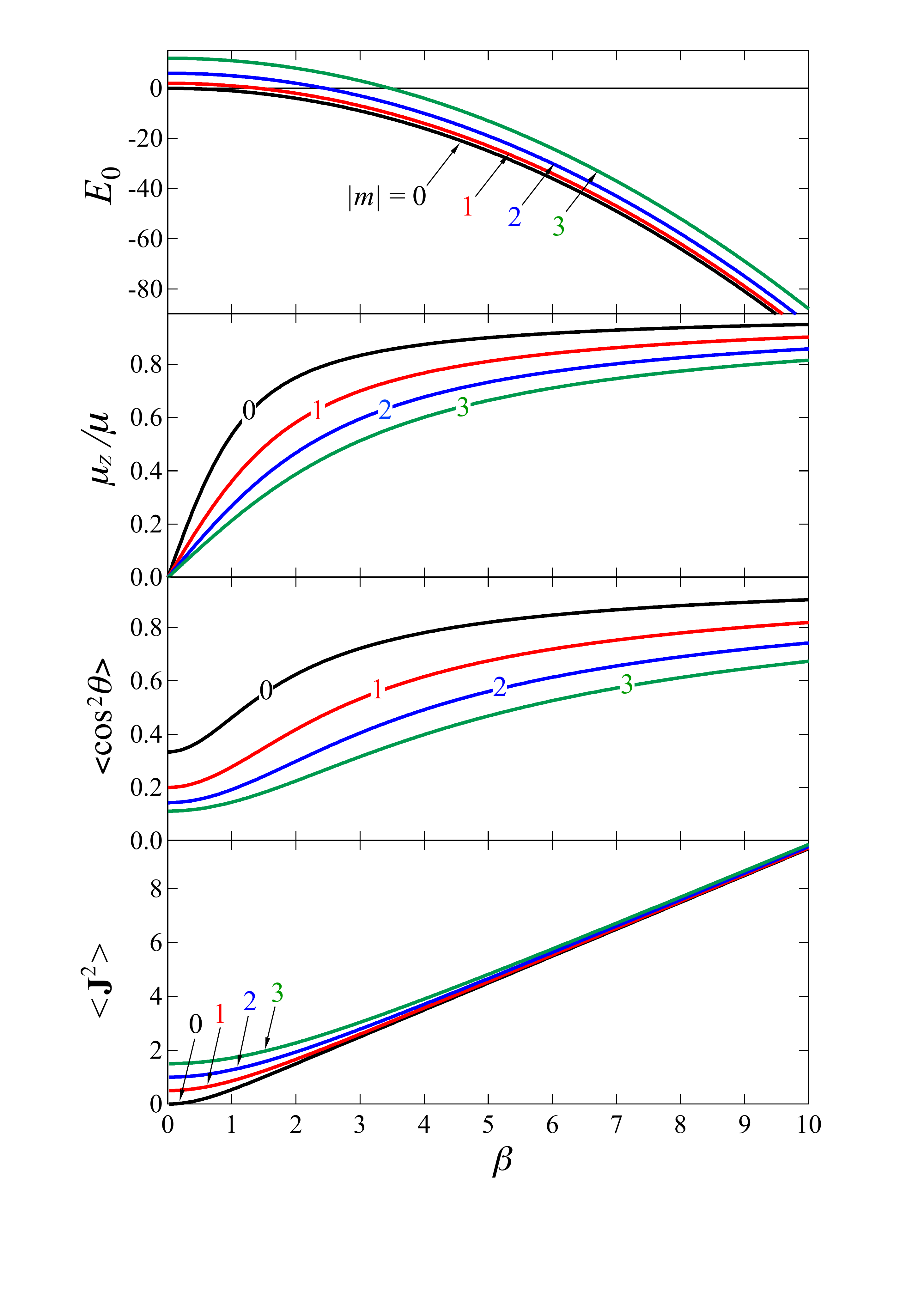}
\caption{\label{fig:mu_Z} Ground-state energies $E_0$ (in units of the rotational constant $B$), space fixed dipole moments $\mu_Z/\mu \equiv \langle \cos \theta \rangle$, alignment cosines $\langle \cos^2 \theta \rangle$, and expectation values of the angular momentum $\langle \mathbf{J}^2 \rangle$ for different $|\tilde{J}=m,m;\beta \rangle$ states as a function of the interaction parameter $\beta$.}
\end{figure}

Exact solutions for molecules in combined fields allow to derive molecular properties analytically. The space fixed dipole moment, $\mu_Z$, is given by the orientation cosine, $\langle \cos \theta \rangle = \langle \psi (\theta) \vert \cos \theta \vert \psi (\theta) \rangle $, and for the exact wavefunction of eq.~(\ref{fviaW}) can be evaluated in closed form,
\begin{equation}
	\label{DipoleMom}
	\mu_Z / \mu \equiv  \langle \cos \theta \rangle  = \frac{ I_{m+3/2} (2 \beta)} {I_{m+1/2} (2 \beta)},
\end{equation}
where $I_n (z)$ is a modified Bessel function of the first kind~\cite{AbramowitzStegun}. 
Fig.~\ref{fig:mu_Z} shows spaced fixed dipole moments corresponding to the states $|\tilde{J}=m,m;\beta \rangle$ for several values of  $m$ as a function of the $\beta$ parameter. The value of $\mu_Z$ rapidly increases with $\beta$. For instance, for $m=0$,  it rises from only $0.54 \mu$ at $\beta=1$ to $0.83 \mu$ at $\beta=3$. In the case of the much  studied $^{40}$K$^{87}$Rb molecule, which possesses a dipole moment $\mu=0.589$ Debye and a polarizability anisotropy  $\Delta \alpha = 54.21$~\AA$^3$~\cite{AymarDulieuJCP05, DeiglmayrDulieuJCP08}, relatively weak fields of $\varepsilon=38$ kV/cm and $I=1.75 \cdot 10^9$ W/cm$^2$ (corresponding to $\beta=5$) give rise to a strongly oriented ground state with $\mu_Z = 0.9 \mu$. This value of $\langle \cos \theta \rangle$ corresponds to the molecular axis confined to librate within $\pm26^\circ$ about the common direction of the fields.

The alignment cosine, $\langle \cos^2 \theta \rangle = \langle \psi (\theta) \vert \cos^2 \theta \vert \psi (\theta) \rangle $, characterizes the molecule's alignment along the $Z$ axis and takes the analytic form,
\begin{equation}
	\label{cos2}
	\langle \cos^2 \theta \rangle  = \frac{2\beta^2 {}_0\tilde{F}_1 (; m+7/2; \beta^2) + {}_0\tilde{F}_1 (; m+5/2; \beta^2) }{2 {}_0\tilde{F}_1 (; m+3/2; \beta^2)},
\end{equation}
with ${}_0\tilde{F}_1 (; a; z) = {}_0F_1(; a; z)/\Gamma(a)$ a regularized confluent hypergeometric function~\cite{AbramowitzStegun}. Fig.~\ref{fig:mu_Z} shows $\langle \cos^2 \theta \rangle$  of the states $|\tilde{J}=m,m;\beta \rangle$ for several values of  $m$ as a function of the $\beta$ parameter. The $|0,0;\beta \rangle$ state exhibits quite a strong alignment with the alignment cosine rapidly approaching with increasing $\beta$ the value of 0.8, which corresponds to a libration of the molecular axis about the polarization vector of the radiative field with an angular amplitude of $27^\circ$.

The expectation value of the angular momentum is related to the orientation cosine, eq.~(\ref{DipoleMom}), via
\begin{equation}
	\label{J2}
	 \langle  \mathbf{J}^2  \rangle  = \frac{m}{2} + \beta \frac{I_{m+3/2} (2 \beta)} {I_{m+1/2} (2 \beta)} \equiv \frac{m}{2} + \beta \langle \cos \theta \rangle,
\end{equation}
We note that the dependence of $\langle  \mathbf{J}^2  \rangle$, shown in Fig.~\ref{fig:mu_Z}, becomes asymptotically linear in $\beta$ for all the values $|m|$, cf. eq.~(\ref{J2}).

By making use of eqs.~(\ref{SEcombinedfields}) and (\ref{OmegaDeltaOmega}), one can show that the eigenenergy becomes:
\begin{equation}
	\label{ExpectValuesCond}
	E_0 \equiv \langle  \mathbf{J}^2  \rangle + m^2 \biggl < \frac{1}{\sin^2\theta} \biggr > - 2\beta (m+1) \langle \cos \theta \rangle - \beta^2 \langle \cos^2 \theta \rangle
\end{equation}

Hence by invoking supersymmetry, we found a condition,  $\Delta \omega = \frac{\omega^2}{4 (m+1)^2 }$, under which the 3D molecular Stark effect problem for a polar and polarizable molecule subject to to a combination of collinear nonresonant electric fields becomes exactly solvable for the  $|\tilde{J}=m,m;\omega,\Delta \omega\rangle$ family of ``stretched'' states. We also demonstrated that SUSY factorization enables constructing families of isospectral potentials, realizable with combined fields. 

Interestingly, it is possible to glean the reason as to why the exact solution of eq. (\ref{SEtransformed}) is obtained for only one particular relation between the field strength parameters, eq. (\ref{OmegaDeltaOmega}), from the semiclassical (WKB) approximation~\cite{FriedrichTrostPhysRep04}. The eigenfunction of a 1D Schr\"odinger equation assumes the WKB form $f(\theta) \propto \exp [i S(\theta)]$, with $S(\theta)$ the action of the underlying classical system. For the ground state of the potential~(\ref{Vminus1D}), the action satisfies the differential equation, $S'(\theta)^2 - i S''(\theta) = V_-^\text{(1D)}$, whose solutions are obtained by expanding $S(\theta)$ in powers of $\hbar$. It turns out that in the case of the combined field strengths connected via eq.~(\ref{OmegaDeltaOmega}), the series converges to the following exact expression,
\begin{equation}
	\label{Action}
	S (\theta) =  \frac{1}{2 i} \left[ 2 \beta \cos \theta + (2m+1) \ln (\sin \theta) \right],
\end{equation}
which, when substituted into $f(\theta)$, yields the exact, closed form wavefunction, eq. (\ref{fviaW}).

We note that the exact $|0,0; \beta \rangle$ wavefunction can be also obtained as a ``curious eigenproperty'' by the method outlined by von Neumann and Wigner in 1929. They showed that by imposing the integrability condition on the sought wavefunction, a class of potentials could be derived that support a localized bound state embedded in the continuum ~\cite{vonNeumannWigner29, StillingerHerrickPRA75}.

The analytic expressions for the characteristics of the strongly oriented and aligned states provide a direct access to the values of the interaction parameters required for creating such states in the laboratory. Moreover, the available analytic eigenproperties could serve to simplify and, simultaneously, render more accurate, models of  many-body systems subject to electric fields, a common scenario for, e.g., ultracold polar gases~\cite{BaranovPhysRep08}.

Our special thanks are due to Gerard Meijer for encouragement and support. One of us (S.K.) thanks the ARO for financial support.

\bibliography{References_library}
\clearpage
\end{document}